\documentclass[sigconf]{acmart}

\usepackage{booktabs} 
\usepackage{comment}
\graphicspath{{./img/}{./plot/}}
\DeclareGraphicsExtensions{.pdf,.jpeg,.png}
\usepackage{amsmath}
\usepackage{listings}
\usepackage{blindtext}
\usepackage{units}

\setcopyright{none}

\acmConference[HIP3ES'18]{High Performance Energy Efficient Embedded Systems Workshop}{January 2018}{Manchester, United Kingdom} 
\acmYear{2018}
\copyrightyear{2018}

\begin{document}
\title[Development of Energy Models for Design Space Exploration]{Development of Energy Models for Design Space Exploration of Embedded Many-Core Systems}

\def\wu{\texorpdfstring{\textsuperscript{*}}{}}
\def\wg{\texorpdfstring{\textsuperscript{\dag}}{}}
\author{
	Christian Klarhorst\wg, Martin Flasskamp\wg, Johannes Ax\wg,
    \texorpdfstring{\\}{} Thorsten Jungeblut\wg, Wayne Kelly\wu, Mario Porrmann\wg, and Ulrich R\"uckert\wg
}
\affiliation{	\begin{tabular}{ccc}
		{\wg\ }Cognitronics and Sensor Systems Group,	& & {\wu\ }Science and Engineering Faculty \\
		CITEC, Bielefeld University, 					& & Queensland University of Technology \\
		Bielefeld, Germany 								& & Brisbane, Australia \\
		\email{cklarhor@cit-ec.uni-bielefeld.de}		& & \email{w.kelly@qut.edu.au} \\
	\end{tabular}
}

\renewcommand{\shortauthors}{C. Klarhorst et al.}

\begin{abstract}
This paper introduces a methodology to develop energy models for the design space exploration of embedded many-core systems.
The design process of such systems can benefit from sophisticated models.
Software and hardware can be specifically optimized  based on comprehensive knowledge about application scenario and hardware behavior.
The contribution of our work is an automated framework to estimate the energy consumption at an arbitrary abstraction level without the need to provide further information about the system.
We validated our framework with the configurable many-core system CoreVA-MPSoC.
Compared to a simulation of the CoreVA-MPSoC on gate level in a \unit[28]{nm} FD-SOI standard cell technology, our framework shows an average estimation error of about 4\,\%.

\end{abstract}


\maketitle

\section{Introduction}
Embedded system designers often have to deal with trade-offs.
One is to optimize a system that performs acceptable in the general case instead of building an application-specific system.
A general purpose CPU excels an ASIC in flexibility but underlies it in performance.
However, a high configurability of hardware and software increases the available design space tremendously.

An embedded system passes through many different design stages, involving different groups of developers from the hardware, software and system design domain.
All have their own perspective on the system at different abstraction levels.
For instance a software developer does not know in detail how the usage of different hardware components affects the performance or energy consumption of an application.
Just as a hardware developer may not have enough information about specific application requirements in an early design stage. 
Each group relies on estimations of specific system metrics that are gathered by tools to perform design decisions.
Building optimized designs for specific use cases often requires an information flow between all involved groups in order to find solutions that are holistically optimal, which leads to a hardware software co-developing process. 
Therefore, it is appropriate to postpone trade-off decisions during the design process in order to decide at the right time, as informed as possible and for the right use case scenario.
However, keeping more degrees of freedom lead to a growing design space on all different abstraction levels and the available and explored space better has to be handled automatically.

\begin{figure}[b]
\centering
\includegraphics[width=.85\linewidth]{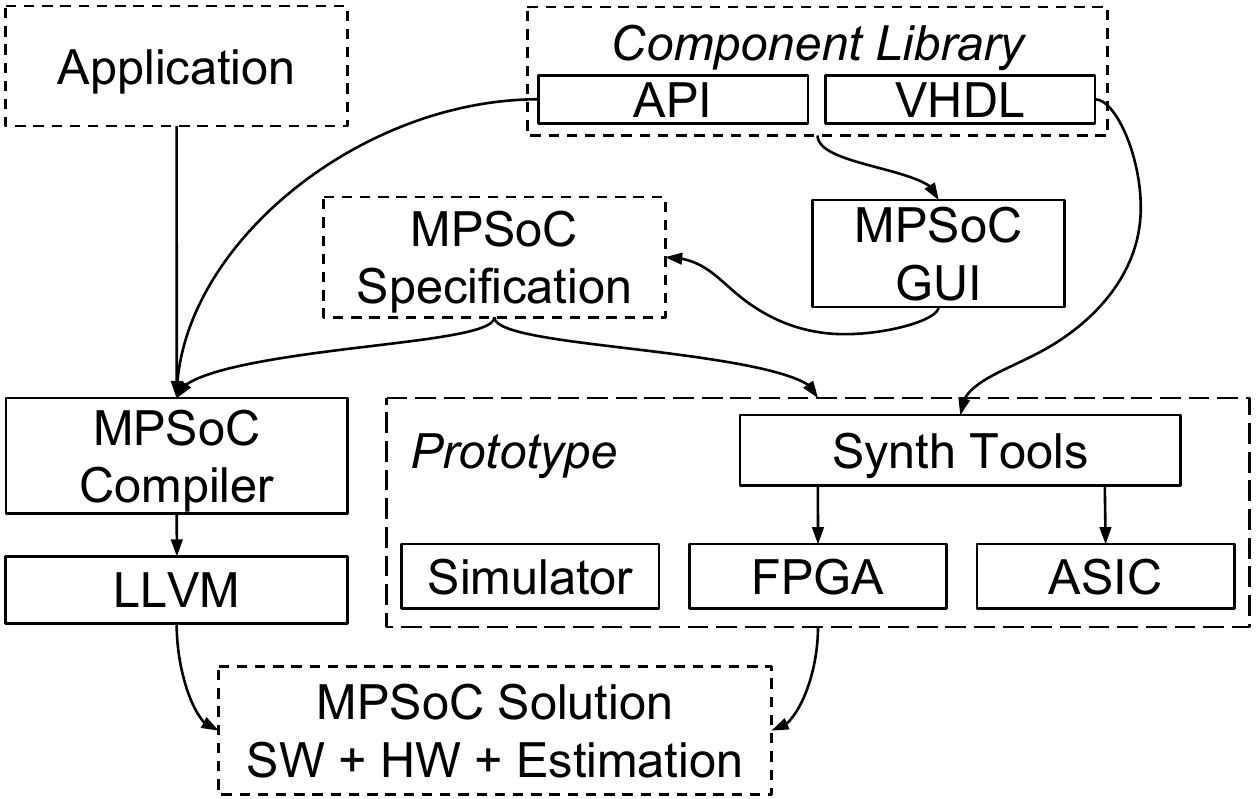}
\caption{Interaction of CoreVA-MPSoC toolchain components}
\label{fig:coreva_toolchain}
\end{figure}

Our target platform is the embedded MPSoC architecture CoreVA-MPSoC, which is highly configurable at design time.
The CoreVA-MPSoC toolchain has the objective to create an interface between the different developer groups.
It consists of an MPSoC system specification in XML, parameterized VHDL SoC components, a graphical user interface to configure MPSoCs, a cycle accurate system simulator, a software backend based on the LLVM compiler framework and our CoreVA-MPSoC compiler together with a broad range of single and many-core applications.
The VHDL SoC components are bundled with their API specification into a component library.
The interaction between these components is depicted in Figure~\ref{fig:coreva_toolchain}.

The system needs to be accurately modeled during the design process of hardware and software despite of the high configurability.
This can be achieved by combining partial models of individual components like execution units or memories to a complete system model.
There are many different metrics to evaluate the system depending on the specific abstraction level.
In \cite{Flasskamp2016} we presented a model to estimate the throughput of an MPSoC.
In this work we focus on modeling the energy consumption and how these models influence the whole MPSoC design process.
Energy efficiency is a crucial design aspect for all kinds of processor systems.
Obviously, battery powered embedded systems are limited regarding their run time.
Additionally, the power consumption of high performance systems limits their performance due to heat dissipation.
The power consumption can be addressed at all design stages.
On the hardware development side by choosing or dimensioning components and on the software side by energy-aware compiler optimization. 
Hence, we introduce an energy estimation framework for the CoreVA-MPSoC toolchain.

Current research typically focuses on modeling effects like cache behavior~\cite{Li2009} or components like NoC routers~\cite{Wolkotte2005} of a specific system.
In contrast, we want to cover a broad range of systems as for instance the CoreVA-MPSoC as a highly configurable architecture.
A huge design space implies to trade off the required accuracy against the setup time of the model.
Therefore, our framework generalizes the model definition and automates its creation.
It is able to create an energy model based on the needs of different user groups that variate in granularity, accuracy and available time.

This work is organized as follows.
The hardware and software parts of the CoreVA-MPSoC project are introduced in Sections~\ref{sec:coreva-mpsoc-arch} and \ref{sec:compiler-infrastructure}.
An overview of related work in the domain of modeling and estimating energy consumption is presented in Section~\ref{sec:related-work}.
Section~\ref{sec:method} describes how energy estimation correlates with system state transitions
and Section~\ref{sec:framework} goes into details how the model creation is automatically done by our energy estimation framework.
The analysis of our energy estimation framework is presented in Section~\ref{sec:results}.

\section{The CoreVA-MPSoC Project}
\label{sec:coreva-mpsoc}
As a target platform for our estimation framework presented in this work, we use the self-developed many-core system CoreVA-MPSoC.
The CoreVA-MPSoC is used for application domains like signal processing, vision processing, and encryption in embedded and energy-limited systems.
In \cite{Sievers2017} the CoreVA-MPSoC is presented as a platform for Software-Defined-Radio applications.

\subsection{Hardware Architecture}
\label{sec:coreva-mpsoc-arch}

\begin{figure}[b]
\centering
\includegraphics[width=1.0\linewidth]{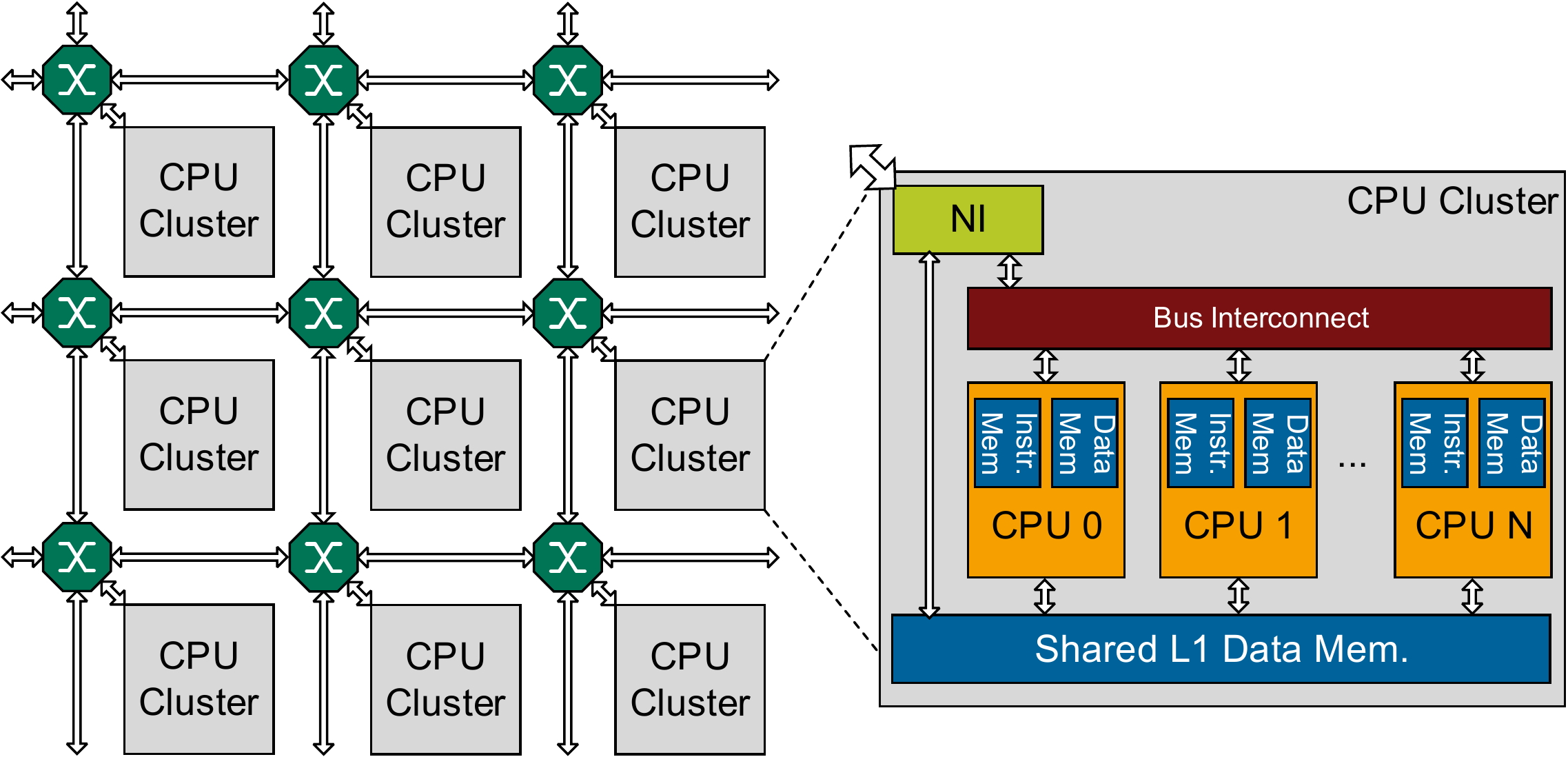}
\caption{Hierarchical CoreVA-MPSoC architecture}
\label{fig:coreva_mpsoc_arch}
\end{figure}

The CoreVA-MPSoC can be classified into the domain of many-core architectures. 
Many-core architectures are defined by a very high number of small-sized CPU cores, which are able to run applications with high resource efficiency.
Kalray's MPPA-256~\cite{DeDinechin2013} and Adapteva's Epiphany are typical examples for such many-core architectures.
In contrast to the Epiphany, our CoreVA-MPSoC features a hierarchical communication infrastructure.
The hierarchical infrastructure consists of a Network-on-Chip (NoC) interconnect that couples several CPU clusters (cf.~Figure~\ref{fig:coreva_mpsoc_arch}).
Within each CPU cluster several VLIW CPU cores are tightly coupled via a bus-based interconnect.

The hardware architecture of our CoreVA-MPSoC is highly configurable at design time on all different hierarchy levels.

Basic building block and lowest hierarchy level of our MPSoC is the VLIW CPU CoreVA that is designed to provide high resource efficiency~\cite{Lutkemeier2013}.
According to the application domain, the number of VLIW slots, ALUs, multiplication, and division units can be configured at design time.
Each CPU has local instruction and data memories, which are used as software-managed scratchpad memories, without the use of caches.

A CPU cluster is used to tightly couple up to 32 of those CPU cores using a bus-based interconnect.
The number of CPUs and the interconnect fabric can be configured at design time and supports either AMBA AXI4 or Wishbone standard, each featuring a crossbar or a shared bus~\cite{Sievers2015b}.
Within a cluster, a CPU can access the local data memory of every other CPU in a Non-Uniform Memory Access (NUMA) fashion.
Flexibility and communication bandwidth can further be improved by integrating a shared data memory into a cluster, which is tightly coupled to each CPU of the same cluster~\cite{Sievers2015}.

For realizing MPSoCs with dozens or hundreds of CPU cores a Network on Chip (NoC) is introduced to the CoreVA-MPSoC.
The NoC acts as a second interconnection hierarchy level and connects multiple CPU clusters.
The NoC features a packet based wormhole switching.
Packets are segmented into flits with 64\,bit payload data.
The NoC interconnect is built up of routers, each having a configurable number of ports. 
This flexibility permits the implementation of most common network topologies.
Each CPU cluster is connected to the NoC via a network~interface~(NI)~\cite{Ax2015}, which bridges between the address-based communication of the clusters and the flow- and packet-based communication of the NoC. 
Therefore, it provides a certain number of independent uni-directional channels. 
These channels are synchronized and the packet data is directly stored to and read from each CPU's local data memory or the cluster's shared L1 data memory.
Hence, the CPUs can benefit from the low access latency of its local memory.

\subsection{Programming Infrastructure}
\label{sec:compiler-infrastructure}

To program the MPSoC, the inter-CPU communication is encapsulated by our communication C-library (API), transparently for both cluster and NoC communication. 
The CPUs communicate via a block-based synchronization model and unidirectional communication channels~\cite{Kelly2014}.
To run C applications, we have created a C compiler tool chain based on the LLVM compiler infrastructure.
Our custom backend supports VLIW and SIMD vectorization.

Nevertheless, efficiently programming a complete MPSoC with many CPUs is a challenge for the programmer.
For this reason we have developed a parallelizing compiler for streaming applications to assist in programming the CoreVA-MPSoC~\cite{Kelly2014, Flasskamp2016}.
This compiler, called CoreVA-MPSoC compiler, processes applications written in the StreamIt language~\cite{Thies2002}.
An application is represented by a structured data flow graph, which describes the inherent parallelism of its tasks.
Our compiler for streaming applications searches for a valid partition with the best performance.
It utilizes an approach based on simulated annealing to map the tasks of a program onto the individual CPUs of the MPSoC.
A typical optimization goal is to find a placement for the application's tasks which maximizes the throughput or minimizes the latency of an application.
During the partitioning process the compiler exploits three degrees of freedom to alter an application's data flow graph.
Firstly, the compiler decides on which processor a task is placed.
Secondly, a task can be cloned to exploit data parallelism.
Thirdly, the granularity of work done in each iteration can be increased to reduce the overhead of communication.
These changes are called mutations and further increase the search space for the partitioning algorithm.

Every partition is judged regarding the achieved performance, by a Simulation Based Estimation (SBE) model, which combines a single execution-based simulation and an analytic approach~\cite{Flasskamp2016}.
MPSoC configurations with the same number of CPUs can differ in many ways like topology, communication infrastructure and memory architecture.
All these characteristics are modeled in our SBE to achieve accurate performance estimations.
In addition, a partition is also checked if any hardware limits of a specified MPSoC configuration are exceeded.
Therefore, our SBE model estimates the consumption of data memory and enables the partitioning algorithm to reject partitions exceeding the physically available memory.

Finally, the best partition is used to generate an individual C code file for each CPU of the MPSoC.
With the energy estimation framework presented in this paper, our SBE model can be extended to estimate the expected energy consumption of an application and perform energy-aware compilers optimization.

\section{Related Work}
\label{sec:related-work}
Energy consumption is still a research topic of much interest.
On the one hand power consumption imposes limits on the system performance due to e.g. heat dissipation.
On the other hand power efficiency often limits the diversity of target applications. Research focuses on energy models, simulators and configuration space exploration, often to characterize various system aspects.
Differences are in the purpose, target user group, level of detail and the method to infer and analyze those models.
The purpose might be mainly driven by the needs of the various development process levels to provide benefits for the involved user groups.
Research was often used to limit the design-space and point out design inefficiencies as early as possible~\cite{Vasilakis2015}.

In the area of hardware development research focuses on different levels of detail.
On gate level Rosell\'o et~al.~\cite{Rossello2005} present an energy model for domino CMOS gates based on a detailed description of internal capacitance switching and discharging currents. Many models target specific processor components for example Dramsim~\cite{Dramsim2} and Cacti~\cite{Cacti} are well-known tools used for timing and energy analysis of DRAMS and caches.
In~\cite{Wolkotte2005} Wolkotte et~al. focus on NoC routers while the Orion2~\cite{ORION2} network simulator models complete Network on Chips.
Energy simulation tools for different kinds of processors are provided by the frameworks Watch~\cite{ISCA00} for CPUs and GPUWatch~\cite{ISCA13} for GPUs.
Just as by the work of Vasilakis et~al.~\cite{Vasilakis2015} for ARM processors and Jordans et~al.~\cite{Jordans13} for VLIW processors.
An instruction-level energy characterization of the Adapteva Epiphany SoC was done by Ortiz et al. in \cite{Ortiz2017}.
They measured the energy consumption of the SoC with a digital current meter while executing a number of microbenchmarks.
The McPat framework~\cite{Li2009} focuses on complete SoCs by describing fundamental components of multi-core and many-core processor configurations in a flexible XML interface.
Specific properties and implementation details like cache behavior and timing are described quantitatively.
By evaluating the model McPat estimates power, area and timing.
Hesse et al. created a power consumption model of a complete circuit board of a wireless sensor node~\cite{Hesse2016}.
The model containing sensor, microcontroller, wireless transceiver and power supply is utilized for choosing components at an early design stage.
In the compiler research area the configuration space of compiler options and its impact on power consumption was explored by Pallister et~al.~\cite{Pallister2013}.
They performed power measurements on five embedded platforms to ensure all architectural effects are captured.
All user groups focus on different levels of detail or granularity and admit different error margins.

Besides the target platform or components the aforementioned work further differs in their methodology.
The McPat framework~\cite{Li2009} estimates energy based on a model specification while other work gathers data from various sources to fit models like Ortiz et~al. did by measuring the MPSoC~\cite{Ortiz2017}.
Multiple options to create and gather data for the second option may be found in research.
Microbenchmarks which are often handwritten are used for the behavior creation in contrast to real applications or benchmark suits. Furthermore data can be gathered by simulating those systems or by measuring final ASICs under real world conditions.
The simulation method provides fine granular power analysis, e.g. it is possible to gather power data for all components of a CPU using gate level analysis at the cost of an often compute intensive but scalable simulation 
Finally, the simulation method is able to provide data before tape out and it is able to simulate various environment conditions. While the measurement method provides better accuracy because it incorporates all physical effects and uses real word conditions.

\begin{figure}[b]
\centering
\includegraphics[width=.85\linewidth]{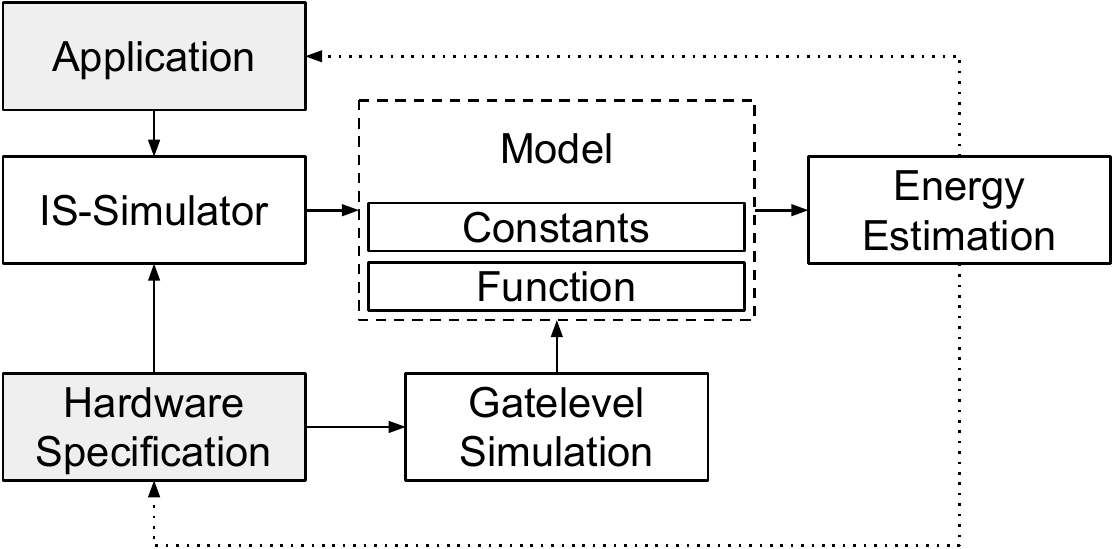}
\caption{Exemplary energy estimation of instruction set simulation by model created at gate level}
\label{fig:model_inference}
\end{figure}

\section{Energy Estimation Methodologies}
\label{sec:method}

Energy consumption of digital circuits originates from different physical effects.
The static power consumption is the result of leakage current of the transistors.
While the dynamic power consumption is composed of switching current, a transistor switches from one logic state to another, and the switching frequency dependent load of charging external load capacitance.
The state of the entire system can be described by the states of all components.

Our idea is to describe the system state and correlate energy consumption with the system state transitions.
This also gives us the ability to correlate energy consumption from a lower abstraction level with high-level state transitions.
Figure~\ref{fig:model_inference} shows exemplary how the result of gate level simulations can be utilized to develop an energy model on logic level.
This model can be used together with a cycle accurate system simulator to estimate the energy consumption.

\subsection*{Model Constants and Functions} \label{sec:system-state}
The state of a system can be described at all abstraction levels.
An overview of abstraction levels shows Table~\ref{table:abstraction_layers}.
On circuit level the system state is described by differential equations of the transistors.
Similarly, instructions define the system state at the higher cycle abstraction level.
\begin{table}[b]
\caption{Abstraction levels}
\label{table:abstraction_layers}
\begin{tabular}{l l l l}
Abstraction Level & System State & Granularity \\ \hline
Circuit 	& Transistors	& Differential Equations 	\\ Logic 		& Gates 		& Boolean Functions 	\\ RTL 		& Registers 	& RTL Spec. 	\\ Cycle		& Instructions  & Instruction        \\
Task 		& Functions 	& System Spec. 	\\ Application	& Program 		& Application Spec.	\end{tabular}
\end{table}
Without additional information we could now perform energy measurements and correlate them with state transitions.
A comprehensive model containing all possible state transformations could be created by performing $n_{meas} = n_{states} \times n_{states}$ measurements.
But unfortunately an incalculable amount of ways to describe the system state exist.

\begin{figure}[b]
\centering
\includegraphics[width=1.0\linewidth]{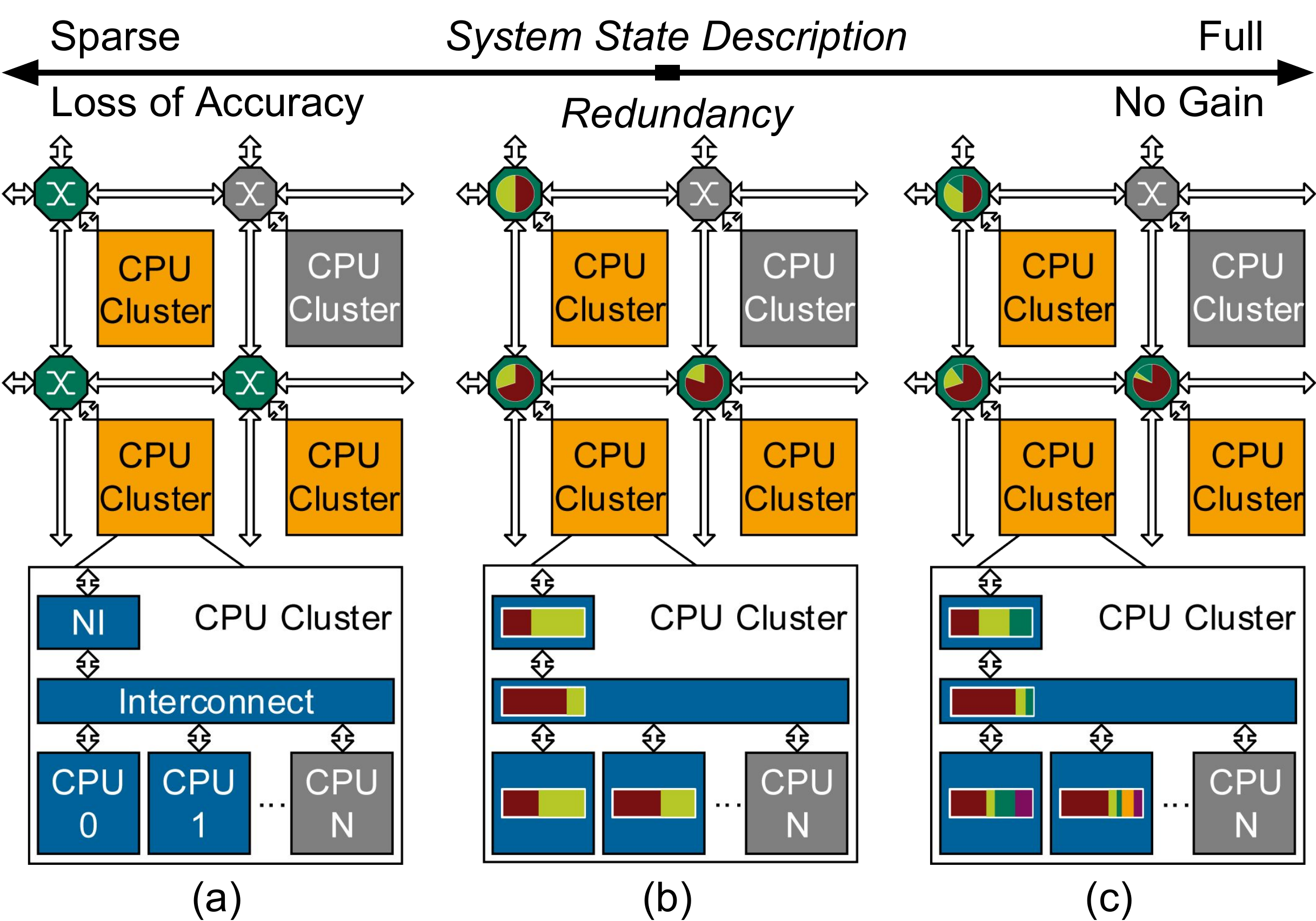}
\caption{Abstraction levels of system state description per period of measurement}
\label{fig:system_state_models}
\end{figure}

This correlation is visualized in Figure~\ref{fig:system_state_models} on the basis of the CoreVA-MPSoC.
Figure~\ref{fig:system_state_models}~(a) show a sparse model which only distinguishes whether a component was used (coloured) or unused (grey) during the period of measurement.
In contrast the model in (c) breaks down the state of each component to a very fine-grained state description, visualized by multicolored bar charts on the CPU components and pie charts on the NoC routers. For instance this model considers the variation of power consumption for different CPU instructions.
A trade-off between both attempts depicts model (b), which considers the time a used components was in active or idle state.
In this case the state description of a component contains no redundant states.

On cycle level system state transitions might be caused by a sequence of instructions as a result of an execution of a function.
The energy consumption per instruction forms the model constants on cycle level (cf.~Figure~\ref{fig:model_inference}).
Setting up the model includes the creation of system behavior, the capturing of the system's energy consumption during simulation and the final model determination.
Accordingly, the setup time for a simple CPU core model would include the generation of micro benchmarks, gate level simulations and fitting the model constants.

On each abstraction level the system state is modeled with a different granularity.
Therefore, we added a function to the model to transform the state description from simulation into the state description used by the model.
It is able to add, change and remove state information and provides the necessary flexibility while it may also include additional system assumptions to simplify e.g. reduce the system description.
This reduction results in faster model evaluation and setup time.
The same model constants can be reused on different abstraction levels by transforming the system state with different model functions.

The model is evaluated by executing the model function and combining the resulting state with the model parameters.
Since no simulation is required the model evaluation is only a fast computation.

\subsection*{Exemplary MPSoC Component Models}
As an example the following paragraph describes the creation of an model for an ideal NoC communication.
The NoC of the CoreVA-MPSoC is organized as a 2D-mesh, described in Section~\ref{sec:coreva-mpsoc-arch}.
A transmission can be characterized by the coordinates of the involved CPUs and the size of the transmitted buffer.
So a naive model would contain all possible permutations of the coordinates and all possible packet sizes resulting in a huge model.

The path on which a packet passes through the NoC can be characterized by a sequence of independent hops (routers).
That way, the foregoing model can be drastically reduced by removing the redundancy of individual links.
The reduced model would then contain all number of hops and possible packet sizes together with a function to define the new distance metric \texttt{number of hops}, for example by $n_{\text{hops}} = \text{manhattan\_dist}(x,y)$.

More knowledge about the hardware could be incorporated into the model to reduce it even further.
For example, one could assume that the energy consumption will increase linearly with the packet size (under the assumption of no congestion).
A model containing all available number of hops together with two constants for the linear functions would be reduced again but possibly at the expense of an increasing estimation error.
More detailed results and further analysis follow in Section~\ref{sec:results}.

\section{Energy Modeling Framework}
\label{sec:framework}
The contribution of our work is a framework to estimate the energy consumption at an arbitrary abstraction level without the need to provide further information about the system.
However, the framework allows for adding additional knowledge to the system to improve estimation accuracy or to accelerate the model setup time.

Back annotated simulations on gate level could be employed to create a model, as long as one is in possession of the RTL (VHDL, Verilog) sources of the system and a process design kit or corresponding IP-Cores.
Afterwards the simulation results can be analyzed by commercial tools like  Cadence\textsuperscript{\textcopyright}~Voltus\texttrademark\ to obtain an energy estimation.
Unfortunately, comprehensive gate level simulations for a complex system like the configurable CoreVA-MPSoC architecture would imply a large number of simulations.
Accordingly, these require a lot of simulation time and resources as well as tool licenses.
Therefore, our framework enables the developer to obtain an energy estimation faster and with an acceptable loss of accuracy with the help of reducing the number of necessary simulations by including system knowledge into the model.

We require a set of system states and energy information to find the correlation parameters for the specific energy model. 
The system states (the system behavior) are created by a previously selected application or by microbenchmarks and a gate level simulation.
This simulation provides the necessary state information together with the occurred switching activity data that is fed into the tools for energy measurement.
We further provide a trace infrastructure, implemented up to the C level, in order to provide a method to create additional, high level system state information that could be used in the model or by the model function.
We developed a tool to automatically create a broad range of microbenchmarks based on user preferences and on the available information in the toolchain.
The tool is able to variate the use pattern of all SoC components that have a specified API description.
This includes for example all different communication and memory components.
It is further able to create microbenchmarks for the CoreVA CPU cores.
This is done automatically by parsing the instruction set description already available in the software toolchain.
The tool always generates setup code in front of every test case in order to bring the CoreVA into a defined state to have a reproductive measuring environment.

\section{Validation} \label{sec:results}

There are many effects in a System on Chip that have an impact on the energy consumption.
This work focuses on a many-core system with a 2D-mesh NoC which connects multiple clusters each containing four fully connected CoreVA CPUs.
The CoreVA CPU cores include two VLIW slots with one load/store and multiplication/division unit.
Each CPU is configured with \unit[16]{kB} of instruction memory organized in two banks and \unit[16]{kB} of local data memory equally organized.
Each cluster additionally contains \unit[64]{kByte} of shared data memory, tightly coupled to each CPU and organized in eight banks.
Each memory bank in the system contains \unit[2]{k} words with \unit[32]{bit} word-size. 
Each network interface (NI) supports 128 synchronous channels with an effective bandwidth of \unit[64]{bit} per clock cycle (cf.~Section~\ref{sec:coreva-mpsoc-arch}).
The cluster interconnects are carried out as crossbars with a width of \unit[64]{bit}.

\begin{figure}[b]
	\centering
	\includegraphics[width=1.\columnwidth]{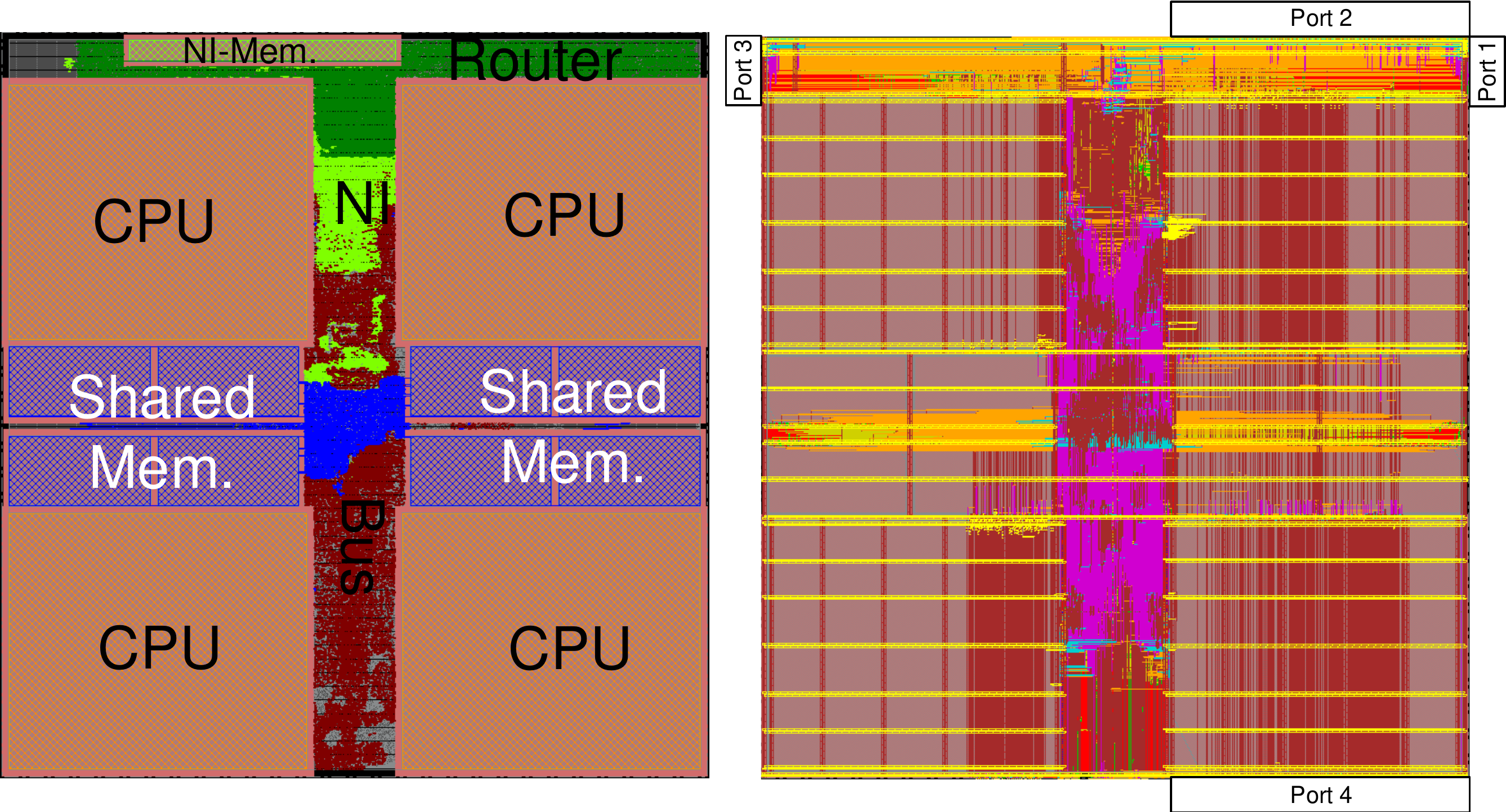}
	\caption{Physical layout of a CPU cluster node. Left: Placed design, Right: Routed design}
	\label{fig:p_a_r} 
\end{figure}

One CPU cluster occupies 0.817\,mm$\mathrm{^2}$ of chip area using a \unit[28]{nm} FD-SOI standard cell technology: STMicroelectronics, 10 metal layer, worst case corner \unit[1.0]{V}, \unit[125]{$^\circ\text{C}$}.
Allows for maximum clock frequency of 700\,MHz.
One CoreVA CPU dissipates 4.5\,mW during idle state and up to 30\,mW during active computation. A full CPU cluster has a total power consumption of up to 150\,mW.
The layout corresponds with the design shown in Figure~\ref{fig:p_a_r}.
In addition to the CPUs, memory and the crossbar, each cluster layout includes NoC components, like the NI and a router with connection ports to all four directions.
Multiple of these clusters can be directly strung together on a top level P\&R step to build a scalable MPSoC with a 2D-Mesh NoC, as we have presented in~\cite{Ax2017}.
As we pointed out earlier, many scenarios exist where accurate energy metrics are important for the final result.
This already begins in an early design stage, e.g. during the specification of the CPU's feature and instruction set.
With our toolchain it is possible to get a fast indicator for the energy consumption of the instruction set.
We automatically create microbenchmarks, which variate just the instructions while setting all registers and the memory content to a specific value.
For instance, the description of CoreVA CPU instruction set (cf. Section~\ref{sec:framework}) contains 155 different definitions of instructions, which results in 20,093 different instruction groups for the two VLIW slots (not all instructions could be placed on an arbitrary slot).
All instruction groups were analyzed for three data patterns. 
\begin{figure}[b]
\centering
\includegraphics[width=0.9\linewidth]{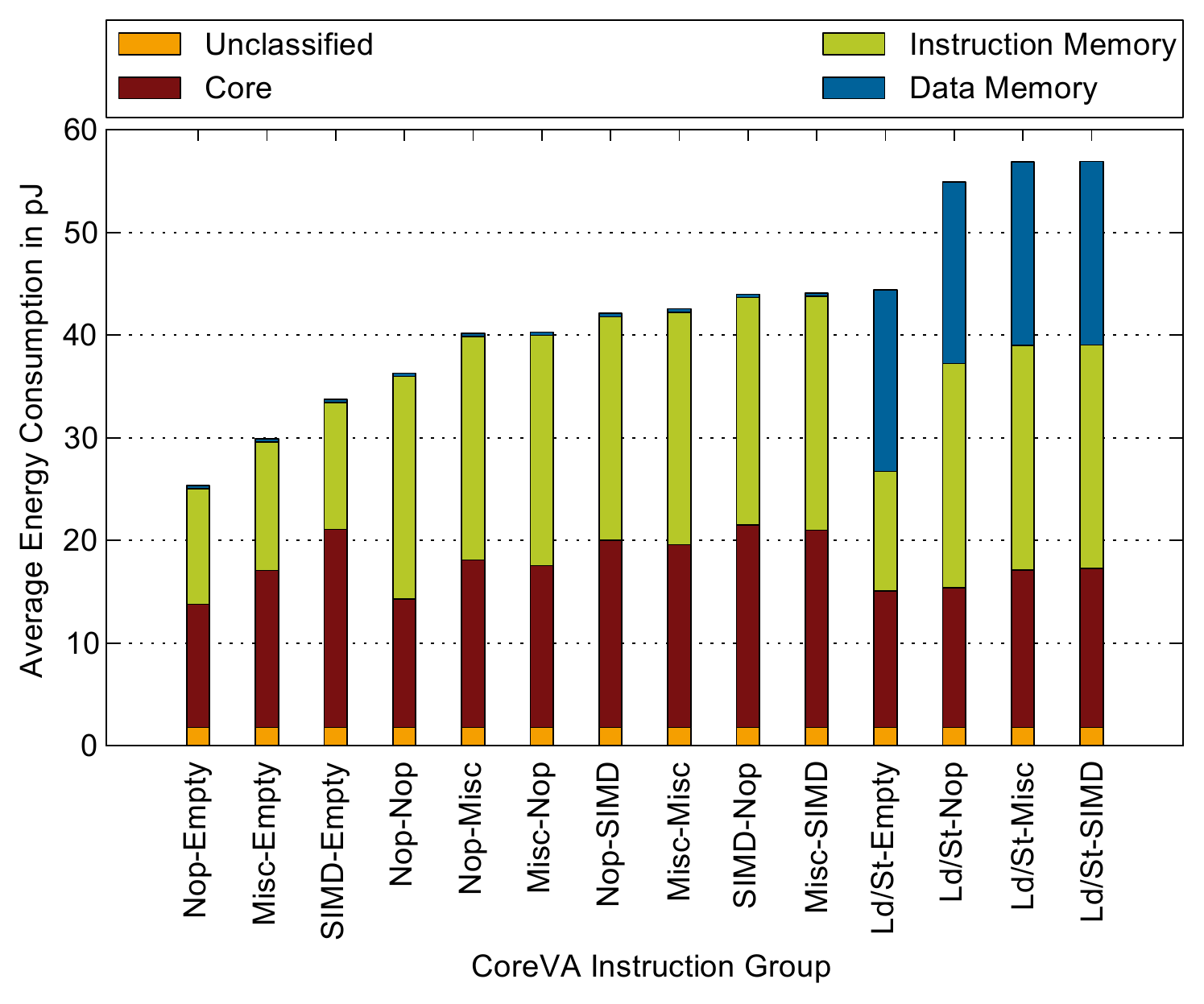}
\caption{Average power consumption of CoreVA CPU instruction groups for two VLIW slots divided into CPU components}
\label{fig:none-all-group}
\end{figure}
This part of the model leads to Figure~\ref{fig:none-all-group} which shows the mean power consumption of the CoreVA CPU instructions for the given configuration.
The energy consumption is shown in pico joule and broken down into the CPU components core, instruction and data memory as well as a small constant unclassified section.
Instructions of similar behavior are categorized into the groups NOP, SIMD, Load/Store and Miscellaneous due to illustrative purposes.
An empty VLIW slot is depicted as {\it Empty}.
Non SIMD arithmetic instructions are included in the Miscellaneous instruction group.

As one might expect, the NOP instruction leads to the lowest energy consumption of the CPU core.
Whereas the most energy is consumed when a SIMD instruction is executed due to the highest core activity.
The data memory shown in blue has obviously an impact only on the power consumption of Load/Store instructions.
The worst case difference between the three data patterns mentioned above is \unit[2.6]{mW}.
The energy consumption of the instruction memory depends on the VLIW instruction format.
A VLIW instruction for a CoreVA CPU with two slots may contain instructions for both or only one slot.
The latter case results in instruction compression to save space in memory and energy on access as seen in Figure~\ref{fig:none-all-group}.
The provided hardware macro of a single CoreVA CPU has a minimum power usage of \unit[15.3]{mW} and a maximum of \unit[38.0]{mW}.
The model creation needs \unit[55]{h} of gate level simulation and \unit[70]{days} of power analysis time to run 60,279 different tests on a single Intel\textregistered~Xeon\textregistered~E5-1650 v4 CPU core.
Actually, the tests are distributed onto 26 CPU cores in parallel.

\begin{figure}[b]	\centering
\includegraphics[width=.9\linewidth]{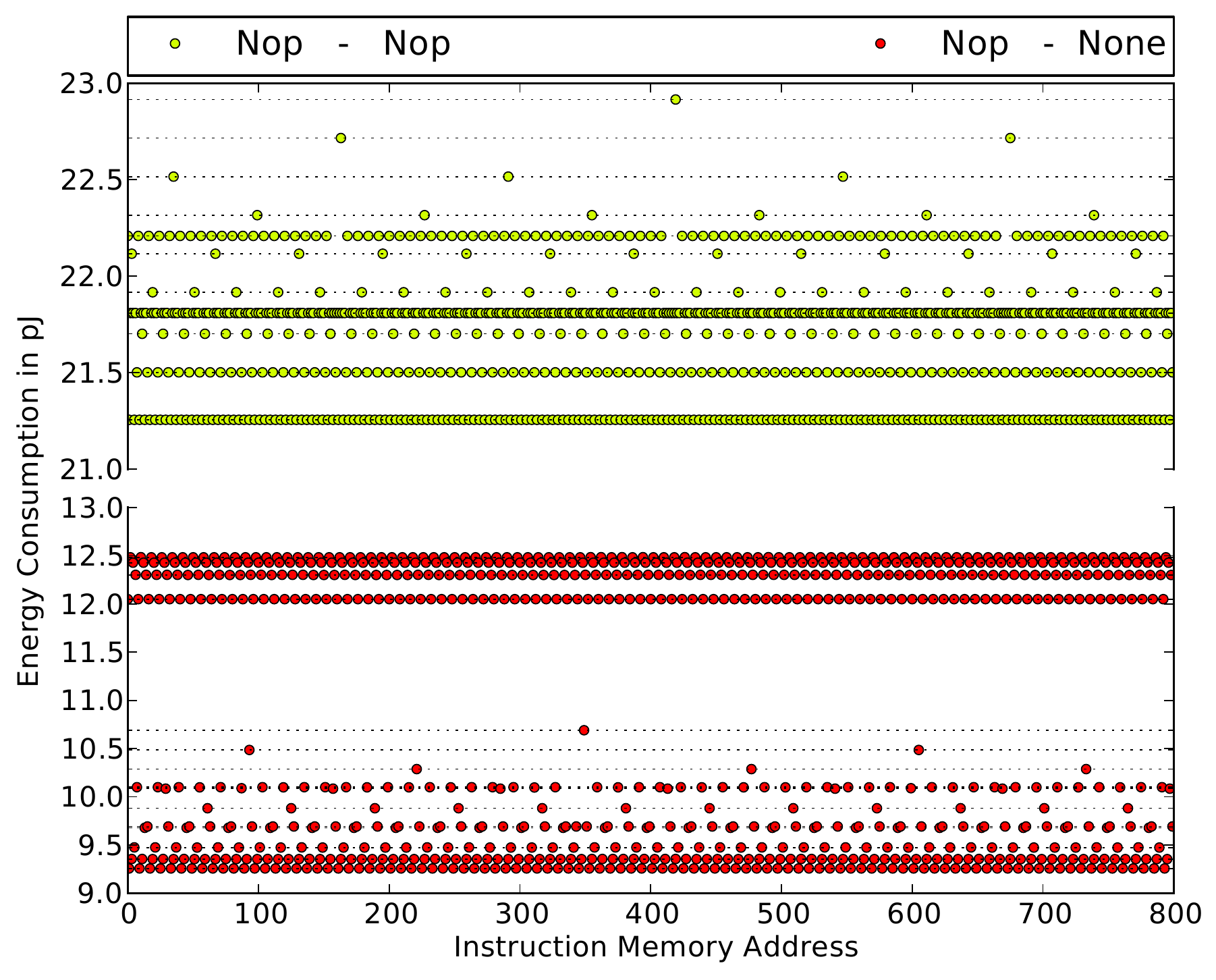}
\caption{Spatial correlation between energy consumption of nop instruction and address in instruction memory}
\label{fig:position}
\end{figure}

Many detailed effects on the power consumption were left aside, like the position of the instruction or the data dependency.
Therefore, we investigated how the specific instruction memory address effects the energy consumption of the instruction memory.
We create microbenchmarks where the position of the instruction is variated while the rest of the system state remains constant.
An \unit[16]{kb} instruction memory is able to store 2048 2-Slot and up to 4096 1-Slot CoreVA instructions because of instruction compression.
The energy consumption for the first 800 instruction memory addresses is shown in Figure~\ref{fig:position}.
All addresses were filled with the same instruction hence the energy consumption only depends on the accessed address and is not a function on the memory content.
The position creates an energy consumption range of \unit[1.7]{pJ} for two VLIW slot instructions and \unit[3.2]{pJ} for a single slot, which can be explained by the internal interconnect of the memory blocks.
We assume that the arrangement of the data points is caused by the tree like structure of the memory address decoder.

\begin{figure}[t]
\centering
\includegraphics[width=.9\linewidth]{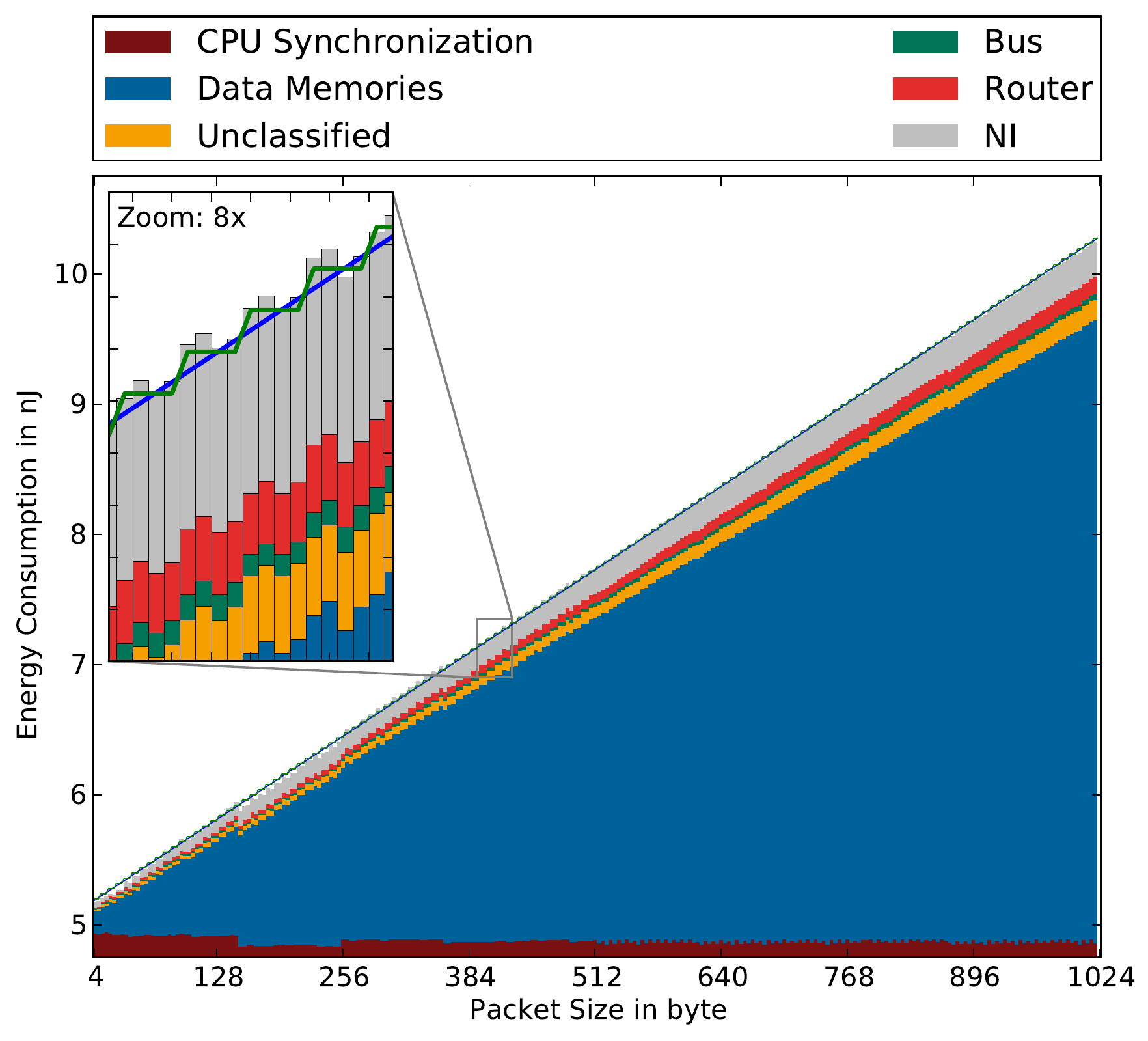} 
\caption{Energy consumption in nJ of NoC transmissions with varying packet size separated into components}
\label{fig:com}
\end{figure}

A communication transfer between two CPU cores across the NoC involves
the two CPU cores itself, routers, the interfaces between CPU cluster and NoC (NI) and other communication infrastructure like the bus interconnect within a CPU cluster.
Our toolchain is able to generate benchmarks with the tool presented in Section~\ref{sec:framework} to analyze communication characteristics in an automated way based on an API description. Therefore, it is possible to variate the packet size for a NoC communication between two cores on different clusters.
Figure~\ref{fig:com} shows the energy consumption of specific transfer sizes.
The packet size variates from 4 to 1024 byte in 4 byte increments.
The zoomed view includes 36 data points from 374 to 446 in 2 byte increments.
Due to the \unit[64]{bit} flit size, into which all NoC packets are segmented (cf.~Section~\ref{sec:coreva-mpsoc-arch}), the plot shows a staircase function.
One bar depicts the energy consumption in nJ and is separated into the shares of the involved components.
CPU synchronization, data memories, bus, router and the NoC interface are shown in red, blue, green, bright red and gray.
The unclassified part shown in yellow contains, among others, especially the parts of the CPU cluster interconnect, which are not assigned to the bus part.
The costs for CPU synchronization are software costs to synchronize the communication channel.
More details about these software costs for communication are presented in~\cite{Ax2015}.

Generating the results of 256 data points for the plot of Figure~\ref{fig:com} took \unit[40]{h} of simulation time and \unit[170]{h} for power analysis.
The data was further approximated with a linear and a staircase function as shown in Section~\ref{sec:method}.
We used 16 data points around the center for the fit in order to reduce the simulation time to \unit[3]{h} and the power analysis time to \unit[11]{h}.
The linear function shown as a blue line had a maximum error of \unit[62]{pJ} while the staircase function shown as a green line had a maximum error of \unit[52]{pJ}.

We validated the applicability of our approach by estimating the energy consumption for a set of five many-core benchmarks from the StreamIt benchmark suite~\cite{Thies2002}.
The analyzed benchmarks are chosen from the signal processing, matrix multiplication and sorting algorithm domain.
Compared to a gate level simulation the average estimation error is about 4\% with a maximum absolute error of 10\% for a simplified model.
The model considers all instruction types and three data pattern but not the instruction position in the memory.
However, the accuracy depends on how many effects are taken into account when establishing the model.

\section{Conclusion}
\label{sec:conclusion}
This paper presents an approach to develop and use energy models for a design space exploration of embedded many-core systems.
The contribution of our work is a framework to estimate the energy consumption at an arbitrary abstraction level without the need to provide further information about the system.
Therefore, we developed an automatic tool flow to fill the energy model with energy information gathered from gate level simulations of basic system states of different components of a many-core system.
Once the energy model is filled, it can be used on higher abstraction levels to get fast and accurate energy estimations for specific applications and hardware configurations of the many-core system. 

To validate our framework, we used our, at design time, highly configurable many-core system CoreVA-MPSoC as a target architecture.
Compared to a gate level simulation of the CoreVA-MPSoC in a \unit[28]{nm} FD-SOI standard cell technology, the average estimation error of our energy model framework is about 4\,\%.
Compared to other approaches, like Ortiz et~al.~\cite{Ortiz2017}, we are able to analyze the energy consumption more fine-grained without the restriction to measure an existing ASIC.
Hence, our model contains internal MPSoC components like instruction memory and VLIW slots as well.
However, Ortiz et~al. achieve a more accurate total energy estimation at instruction level.
The design process of highly configurable embedded systems like the CoreVA-MPSoC benefits from such a sophisticated model.
Applications may be specifically optimized for tailored systems by modeling the different variants and characteristics of system components.
A trade-off can be found based on comprehensive knowledge about application scenario and hardware behavior.
In future work, our MPSoC compiler will use the presented energy model to automatically map streaming applications to our CoreVA-MPSoC, optimized for energy.

\section*{Acknowledgment}
The reported research was supported through project grants KogniHome (German Federal Ministry of Education and Research (BMBF) Grant No. 16SV7054K), the Cluster of Excellence Cognitive Interaction Technology "CITEC" (EXC 277) at Bielefeld University and the BMBF Leading-Edge Cluster "Intelligent Technical Systems OstWestfalenLippe" (it's OWL).

\bibliographystyle{ACM-Reference-Format}
\bibliography{references}

\end{document}